# Scout: Leveraging Large Language Models for Rapid Digital Evidence Discovery

Shariq Murtuza

**Abstract**

Recent technological advancements and the prevalence of technology in day to day activities have caused a major increase in the likelihood of the involvement of digital evidence in more and more legal investigations. Consumer-grade hardware is growing more powerful, with expanding memory and storage sizes and enhanced processor capabilities. Forensics investigators often have to sift through gigabytes of data during an ongoing investigation making the process tedious. Memory forensics, disk analysis all are well supported by state of the art tools that significantly lower the effort required to be put in by a forensic investigator by providing string searches, analyzing images file etc. During the course of the investigation a lot of false positives are identified that need to be lowered. This work presents Scout, a digital forensics framework that performs preliminary evidence processing and prioritizing using large language models. Scout deploys foundational language models to identify relevant artifacts from a large number of potential evidence files (disk images, captured network packets, memory dumps etc.) which would have taken longer to get identified. Scout employs text based large language models can easily process files with textual information. For the forensic analysis of multimedia files like audio, image, video, office documents etc. multimodal models are employed by Scout. Scout was able to identify and realize the evidence file that were of potential interest for the investigator.

Keywords: Digital Forensics; Forensic Analysis; Digital Evidence; Network Packet analysis; Incident Response; Foundational Models; Large language models; Multimodal models.

# 1 Introduction

Digital Forensics is described as the science of identifying then extracting and finally presenting, digital or electronic evidences in the court of law. Digital forensics procedures have been well documented and standardized across Law Enforcement Agencies (LEA). Every country around the globe have strong procedures, methods and laws to tackle cybercrime. The science of digital forensics was invented in the late 80's and early 90's. Since then it has evolved into a multifaceted discipline that involves digital forensics of desktops, laptops, smart terminals, smartphones, remote servers and many more devices. With the increasing number of devices, the software platforms have also evolved with software companies now providing a variety of operating systems. The spectrum of devices one can have has exploded from smart eyewear to wristwatches. Consumer grade computers and laptops have far exceeded anticipation and now come loaded with gigabytes of RAM and terabytes of hard disk storage. In the untimely event of an incident, first responders start by imaging the hard disks and dumping the RAM contents for further analysis. The analysis is becoming more challenging with the increase in the data size.

With growing criminal and civil investigations is leading to major pending issues and induced delays in the due process [1, 2]. The primary reasons for this are the increase in cases involving digital evidences and the increase in the average number of devices that are seized by the LEAs per case. Within the last decade the price per gigabyte (GB) of storage disks and solid state drive (SSD) have came drastically down. This has resulted in larger capacity storage devices being installed as the default configuration in computing devices across all spectrum, mobiles, tablets, desktop computers, laptops etc. [3].

Proper analysis of forensics artifacts is extremely crucial for its acceptability in the court of law. This puts a lot of responsibility on the forensic investigator to identify all the investigation related artifacts from the evidences. Often such a requirement comes with limited time and results in pressure upon the investigator.

Due to the multifold increase in the amount of data seized during investigations, the processing times are also increasing proportionally leading to the calls for research on ontime analysis of seized data. Shaw and Brown [4] discusses in details the pros and cons of the digital forensics triage process along with commenting on the effects of longer processing times for digital seized data. Some of issues they identified included longer trial waiting durations, reduced access to friends and family, job issues due to an ongoing trial among others.

This paper presents an experimental framework Scout, that can be deployed to prioritize evidence processing based on the their potential relevance based on the context of investigation.

A foundational model [7] like the large language models (LLMs) [5] is typically trained on a large corpus of data consisting of different types of text with an ultimate aim to produce human like text. Deep learning architecture is typically used to produce such models [8]. The training corpus typically consists of vast amount of Internet website data such as Common Crawl [6] dataset that consists of billions of crawled web pages.

These models can achieve general purpose language generation and exemplary natural language processing tasks. Creating a LLM involves self supervised and semi supervised learning methods. The primary driving force behind a LLM's capability is the Transformer architecture. This transformer architecture was introduced in the year 2017 by the Google brain team in the paper titled "Attention is All You Need" [9].

The Google brain team introduced the Transformer architecture in that paper [9], which is a novel solution solve to sequence-to-sequence modeling, by using only attention mechanisms instead of using recurrence or convolution. The Transformer architecture has since greatly revolutionized the NLP scenario and has become the de-facto base for many state of the art NLP models such as BERT [10], GPT-3 [11], and others.

Digital forensics is a very important aspect of forensic science as it assists in identification, assessment of and presentation of digital sources in the legal process. It spans a wide range of processes aimed at locating, maintaining, retrieving and interpreting electronic information from different formats and devices. Combining tools including computation, data repairing and exploration, it works towards knowing about the digital world to provide evidence regarding cyber crimes, intellectual theft, fraud among other crimes [12, 13].

During an investigation, the forensic investigator must focus on the identification of artifacts and other digital findings that provides insights into the current investigations based on the context of investigation background. The investigator focuses on the examination and interpretation of electronic data and digital media to reconstruct events, trace digital footprints, and determine the authenticity and integrity of digital evidence. The process encompasses a wide range of digital devices and storage media, including computers, mobile devices, networks, servers, cloud services, and various forms of digital data such as documents, emails, images, videos, and metadata.

The process of digital forensics typically involves several key stages [12, 13, 14]:

1. Identification and preservation: The process of identifying and extracting a potential evidence and maintaining its integrity is the first step. It includes protection of the physical devices, forensic imaging to acquire the data, and creating an unbroken chain of custody over the evidence.

2. Acquisition and analysis: The digital evidence has to be obtained from the provided data sources by following methods that have been approved scientifically and are forensically sound. Investigators use specific tools and techniques to extract and analyze data for pertinent information, for hidden files, erased data and all other elements which could assist in the investigation.

3. Examination and interpretation: Once all the data obtained has been carefully processed and interpreted to create a timeline, the investigator can start to arrange all the pieces of the jigsaw puzzle. This often involves finding the relation between evidences that were obtained. It involves recovering deleted files, examining file attributes, network logs, encrypted data.

4. Reporting and presentation: Once the examination and the interpretation of the identified evidences have been done, the investigative findings and analysis are compiled in a complete report which describes the techniques employed, the materials collected, and the logical reasons for each decision made.

The continuous growth and complexity of technology and cybercrime have resulted in an ever increasing growth of digital forensics. Some of the recent advancements in the field of the digital forensics include the following thrust domains. As cloud computing and storage services become mainstream, investigators are now faced with the task of harvesting and interpreting digital evidence that is stored in the cloud. As a noteworthy development in cloud computing, it is now possible to obtain and interpret data hosted on cloud computing services by Amazon, Microsoft, and Google [15, 16]. Smartphones and smart tablets have now emerged to be an important source of digital information. The focus of developments made in mobile device forensics include obtaining and interpreting data from locked/encrypted mobile devices and obtaining data from messaging and social media applications and backups [17, 18] on Android and iOS based devices. Similarly, the vast of the number and different types of Internet connected devices like the smart appliances, home security systems, wearables, and industrial sensors have also posed unique challenge. IoT forensics essentially entails data extraction and analysis strategies from devices that are connected to the internet or within the Internet of Things ecosystem that have been used to perpetrate a cyber, privacy, or any other crime involving IoT systems[19, 20, 21]

In recent years the use of machine learning and AI in digital forensic process has shown immense potential to improve and fasten the overall investigation process. Large volumes of information can be analyzed and trends and anomalies present would be quickly reported[22]. The use of AI assisted or AI driven technologies can help in the initial sorting out of evidence and its importance, in the organization of files, and the identifying of potentially harmful activities in the clouds within minutes[23]. This work presents and explores the application of different state of the art foundation models (including large language models, vision models etc.) have been deployed( Llama models, Qwen vision models etc.) to preliminarily analyze, process and identify potentially interesting artifacts from the corpus of seized data. These findings are then presented to the forensic investigator for manual verification on priority basis.

Our contributions are as follows:

- This work presents Scout, a prototype digital forensics framework that can process digital evidence and suggest the forensic investigator the order in which files can be processed to get to the evidence faster.

- At no point during the investigation does Scout interferes with the investigation or the investigator's ability to evaluate evidence either manually or with the help of tools, thus removing any threat of contamination of the evidence.

- The Scout framework can be easily deployed in an entirely offline, on-premises environment, or on a central cloud that can be accessed from multiple locations.

- Scout can process evidence in the form of text (emails, office documents etc.), images, videos and captured network packets. Scout's ability can easily be extended by writing plugins for each evidence file type.

- Different foundational models are deployed and the investigator can select between multiple runs of the same model or different models to process an evidence file, allowing different attempts to uncover different facts.

This paper is arranged as follows, Section 1 starts with the introduction and gets the reader acquainted with the background and context of the paper. Section 2 discussed similar advancements done in the academia and the industry. Section 3 sheds light upon the methodology upon which Scout functions while section 4 and 5 presents the results and discussion respectively. This paper is then concluded with section 6.

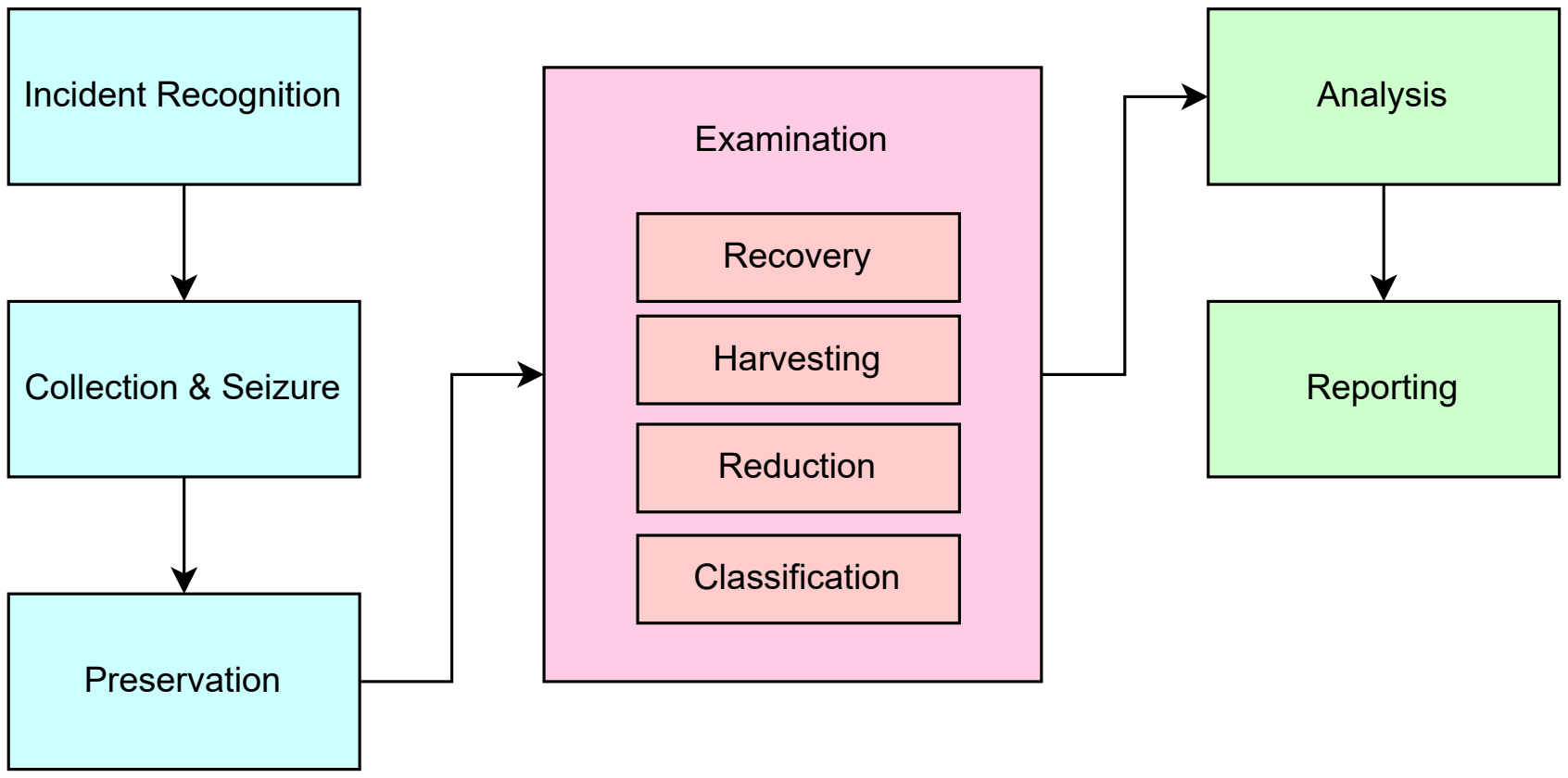

Figure 1: Casey's Traditional Digital Forensic Model

# 2 Related Work

## 2.1 Digital Forensics

A typical digital forensics process involves identification, preservation and later analysis to allow recording the data obtained. This data is obtained usually from electronic devices such as laptop, smart tablets, smartphones, computers etc. The findings identified that were recorded are then produced in the court of law to aid in a criminal investigation or civil litigation. To further standardize the process of these steps, multiple models have been proposed for different scenarios such as smartphone forensics, laptop forensics, online cloud forensics etc. For each of these scenarios the forensic acquisition steps may vary slightly depending on the formats and hardware based data acquisition methods. For mobile forensic acquisition Dhaqm Et al. [24] have performed an exhaustive survey to compare and present the existing forensic process models. Different forensic models are often used to proceed based on mobile operating system, the storage technology being used, the smartphone applications being used etc. Further systematic literature reviews have been done [25] that highlights the common points between the models as well as highlight the ambiguities that often crop up between these models. Javed Et al. [26] have compiled a detailed list of software tools and forensic models that are currently being used.

Casey's model has been widely scrutinized and agreed across the industry. The model, presented in figure 1 is broken into different stages. The first stage is incident recognition that involves the forensic investigation team to identify the actual incident that occurred, identify the equipment involved in the event, recognize the potential sources from where relevant artifacts can be extracted such as hard disks, computers, mobiles, network devices such as switches, routers etc. Once the incident area has been identified, the proper collection and seizure stage shall be conducted to acquire all the evidences. All the computing devices, their associated storage devices, connecting cables, storage devices etc. are to be color tagged and stored in a air-gapped manner typically using a Faraday cage [28]. The preservation of the acquired evidences requires the investigation team to ensure that at no point during the forensic investigation does any of the evidence (seized hardware, acquired files etc.) get tainted due to any intentional or unintentional tampering that may result in the evidence losing its integrity. This is achieved by using cryptographic hashes to check for integrity of acquired files at each step and documenting all the steps that have been taken during the investigation. A log must be maintained to record all the interactions between any evidence and investigating personnel. During the examination stage four major steps are to be taken namely recovery, harvesting, reduction and classification. During recovery forensic investigators try to recover hidden or deleted artifacts that can be of potential interest. Harvesting and reduction involves extracting the identified artifact such as metadata of a file, leftover data in the file system slack etc. and reducing them to get the relevant evidences. Classification requires the investigator to create different classes and sort evidences into them. Once the examination stage finishes, all the evidence is then analyzed and a final investigation report is prepared that will be presented in the court.

## 2.2 Large Language Models

In the last few years Artificial General Intelligence (AGI) [29] and large language models have shown extreme potential in data analysis tasks [30]. Artifacts identification, extraction and processing can be automated using large language models if the models are properly trained and finetuned on evidence extraction. Large language models are text processors and textual artifacts that can be read and understood both with and without context [31, 32]. Often, relevant textual evidence is hidden deep within less important information, causing it to be overlooked by investigators. A large language model on the other hand can easily sift through large amounts of text, summarizing them, identifying relevant bits and contexts.

Current large language models are based on the Transformer architecture built upon tens of million parameters [9]. These models are trained on large amount of textual data. These models have natural language processing capabilities that mimic human understanding like GPT [33], LLaMA [34], Gemini [35], Qwen [36], Qwen-VL [37] etc. ChatGPT [38], which is based on the GPT model series, was the forerunner in introducing the masses to this technology. ChatGPT at the time of it's launch possessed extensively capabilities in the domain of communication, instruction following, task solving among many other complex tasks. The GPT model series functions by processing and compressing textual information as a decoder based transformer model, simultaneously being able to recollect the information it was trained on it. GPT-3 [39], released in the year 2020 was trained on a corpus of size atleast 175B parameters. In November 2022, ChatGPT was released with GPT 3.5 with impressive abilities in performing human instructed tasks. The next iteration in ChatGPT came with GPT 4 in March 2023. GPT 4 has better task solving capabilities while being a multimodal model enabling it to process inputs other than text [40]. This made it far more superior than any other previous version of GPT. In May 2024, GPT 4o (omni model) was released allowing ChatGPT to interact with textual and audio visual inputs in real time. This version of GPT was trained upon data consisting of audio, video along with the traditional textual data [41]. The models developed by OpenAI are not available for custom finetuning and further sharing making them closed models.

One of the biggest player is the LLaMA family of large language models, developed by Meta AI [43]. The family consists of LLaMA (February 2023), LLaMA 2 (July 2023),

LLaMA 3 (April 2024), LLaMA 3.1 (July 2024), LLaMA 3.2 (September 2024) and LLama 3.3 (December 2024) models. With each iteration the LLaMA models have become faster and capable [44].

Docling is a novel tool from IBM [45] intended for processing a document into its constituents. Docling differs from traditional Optical Character Recognition recognizing (OCR). OCR simply ouputs the contents of the document in textual format, whereas Docling outputs a structured JSON or markdown. Docling does uses OCR but only as a part of a document understanding pipeline. Docling uses AI models on each page to identify specific features and content such as layout and table structures. The models have been trained on the DocLayNet dataset [46], and TableFormer [47] for table structure recognition.

## 2.3 Large Language Models and Digital Forensics

Wickramasekara Et al. [48] have conducted a deep analysis of the existing large language models, deep learning techniques and other tools for their suitability in the domain of digital forensics.

Scanlon Et al. [49] explores the applicability of ChatGPT in digital forensic investigation. The study looks into the application of chatGPT into keyword searches, incident responses, scenario building etc. Preliminary findings suggested the application of chatGPT in the domain of digital forensics is questionable since these language models are trained to provide answers to user as a primary goal. These models are not trained for accuracy and precision hence prone to incorrect answers or hallucination. Hallucination defined as ”generated content that is either nonsensical or unfaithful to the provided source content” [50]. Thus if a model produces incorrect or inaccurate result, it should not be considered exceptional, instead hallucinations are quite common for large language models.

The authors suggests using ChatGPT in log analysis, knowledge analysis, generate regular expressions and other scripts. Any use of ChatGPT must be supervised heavily by humans and any output produced must be proofread for logical or other mistakes. This work is based

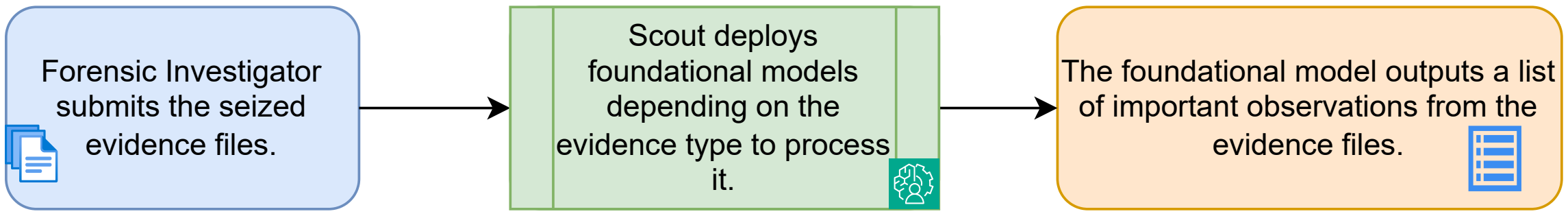

Figure 2: Scout Framework

on the work done by the authors and aims to extend their work.

# 3 Experimental Methodology

This section details the steps that were taken for processing of evidence and analysis using the foundational models to identify files of potential interest. Different seized evidence files requires different models for processing, evidence may have network packets, texts (emails, office documents) etc. are processed using a large language model. Multimedia evidence files such as audio, video and images are processed using vision models or multimodal models that allows their processing. Figure 2 shows the underlying process deployed in our framework Scout.

## 3.1 Network Packets Analysis

For the first round of analysis, the captured network packets were processed and analyzed using different large language models. Direct processing of a PCAP files would be inefficient since the organized structural information in the packet structure would be lost. Thus, before analyzing a PCAP file, its contents are first loaded packet by packet and then individually sent to the large language model for analysis. The accuracy of the analysis obtained is highly dependent on the prompt supplied along with the network packet to the model. An incorrect, ambiguous or unclear prompt will lead to an inefficient analysis. As discussed earlier, large language models work on the principle of probability to predict next tokens based on the current and previous tokens, it can often hallucinate and predict inaccurate and incorrect finding. In the experiments that were conducted different large language models exhibited different levels of accuracy and intelligence while processing the packets. During experiments it was found that smaller models were often incorrect giving wrong results and hallucinating with extra information. It was found that providing additional context of the investigation scenario often helped in obtaining better results.

Llama 3.3 70B [43] parameter model was used for analyzing the network packets. The Llama model was chosen due to its versatility ability and context length of 128K tokens. When compared to the others models such Hermes 3 [51], Llama 3.3 model was found to be producing a more detailed analysis of the network packets.

## 3.2 E-mail data Analysis

Enron e-mail dataset is an often researched and well explored dataset containing real and non-tampered email communication between the employees of Enron Corporation. The dataset consists of around 600K emails containing the email conversation of their 158 employees. The email dataset was released following an investigation by the Federal Energy Regulatory Commission (FERC) after the companies collapsed in December 2001. The Enron dataset is available in multiple formats, as text files, as a database etc. To process the Enron emails using large language models it was expected and also found out that reading and processing emails as a batch yielded better results instead of passing the raw binary files. The investigating prompt was updated with more background information and context explaining the Enron investigation that enabled open ended investigative searches. Large language models are quite adept at summarizing and analyzing text. Further mail analysis was done by summarizing emails and then asking the large language model to identify links between email entities and anomalous behavior. Unfortunately due to the safeguards built into the Llama models while training, the models often understood the digital investigation tasks given as unethical or illegal requests. This was due to their recognition of the evidence files as private emails between individuals and identifying their analysis as invasion of privacy.

This led to the exploration of more obedient models that, even though built with safety measures, still performed the tasks well. One such large language model that was finally selected is Nous research's Hermes 3 model [51]. Hermes 3 models are based upon the Llama 3.1 family of models and have been trained on synthetically generated datasets. Their added advantage over the Llama models is their creativity and reasoning over the Llama models.

## 3.3 Office Documents Analysis

Office documents (.doc, .docx, .odt, .ppt, .pptx, .odp, .xls, .xlsx, .ods, .pdf, .html) files are builtup of markup files, having similar structure like XML with the document contents. Document files were first processed using Docling [45]. The structured information was then passed on to the large language models for analysis. The context (if available) can be passed to along with the prompt to quicken up the processing.

## 3.4 Audio File Analysis

Audio files are analyzed in two separate phases. The first phase involves transcribing the audio file using the state of the art Automatic Speech Recognition model by OpenAI Whisper [52]. Whisper model by OpenAI allows offline speech recognition and translation (between different languages) of audio files. Once the first phase transcribes the audio file using Whisper models, the second phase initiates by processing the transcribed file using Hermes 3 model and raising an alert if any red flag is found. This part is similar to the processing of text files and can be processed similarly.

## 3.5 Image File Analysis

All the previous tasks were requiring the acquiring and analysis of textual and similar information. These tasks can be easily handled and processed using text based large language models. For processing images a multimodal model can be deployed. Multi-modality allows a model to process and analyze audios, image or video based data. Large language models can be used to generate multimedia from text as well as to process multimedia and produce the results in text. A multimodal model harnesses the power of large language models to analyze input data apart from text. Our experiments involved the application of a multimodal large language model, specifically Llama 3.2 [43] and Intel's Llava-llama3 model [53, 54] to identify and analyze the image contents and extract information from them accordingly. Since these vision models are multimodal they can not only analyze the images but also process whether the image had anything related to the case as well.

### 3.6 Video File Analysis

Multimodal models are well capable to process and analyze video files of limited length. Experiments conducted involved passing video files to a multimodal large language model with video processing capabilities. The model provided analysis of videos that can be used to highlight videos of interest from the forensics investigators purpose. Additional context from the investigation in the prompt can also further aid in the identification of videos of interest. For video processing Alibaba's Qwen2 Large Vision Model [55] was deployed. The model is capable of processing videos of around 20+ minutes. Being a multimodal model, Qwen2-VL can also detect and raise red flags based on the investigation context.

## 4 Results

Forensic investigator often have to sift through a long pending list of evidences that require manual analysis. The forensic investigation requires evidence detection, extraction and analyzing for quick evidence processing. This work aims to present a pre-analysis support framework for the forensic investigator to identify files that may potentially be artifacts of interest in the ongoing investigation. This work presents an exploratory analytical study in determining the efficacy of various large language models including processing image and video files using multimodal models. The efficiency of evidence detection depends upon the large language model's capabilities and the resources available at the disposal for the model to run. Time taken for processing files increases with the increase in the file size. The results obtained are discussed below in detail. We decided to hold the release of Scout's accuracy on complete datasets due to the probabilistic nature of foundation models. These model are known to vary in the output produce during different iterations, which may cause certain aspects of their findings being skipped. Once again, it is heavily stressed that Scout is merely a pre-analysis tool meant to facilitate the forensic investigator in deciding the analysis order of the seized evidence.

## 4.1 Network Packet Analysis

This subsection explores the efficacy of the proposed digital forensic tool Scout in processing and extracting relevant information from a PCAP file of captured network packets. Scout was fed the captured packets one by one and ordered to inspect for any point of interest. To further reaffirm the faith in large language models, a PCAP file (containing simple DNS requests) from sample Wireshark captures was taken. This PCAP file was processed without any specific directive given to Scout. Generic instruction to process the PCAP was given to Scout following which Scout not only presented a detailed report but also commented on peculiarities such as identifying repeated DNS requests, multiple ICMP errors as a possible result of the computer being offline. Figure 3 shows the findings of the language model upon the network packets. Packets 113 to 116 were identified are DNS query for *slashdot.org* Scout further identifies the next packets that contained the response of the DNS query (115-119). Scout comments on the presence of only DNS packets, because in normal scenarios other network traffic is also present in the capture.

## 4.2 E-mail Data Analysis

Scout's performance in analyzing and extracting relevant artifacts from email data was successfully demonstrated by analyzing, extracting and summarizing the information from the

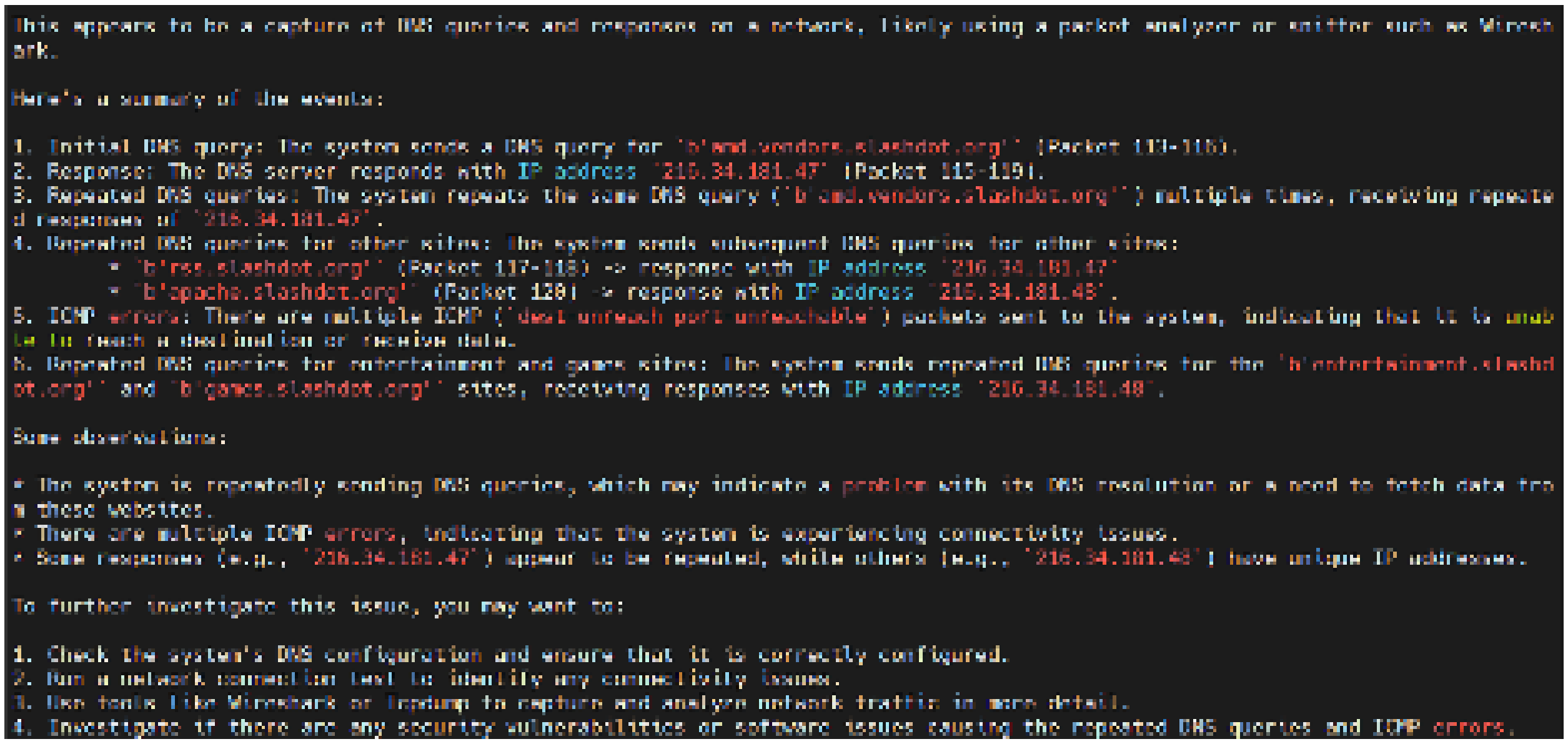
This appears to be a capture of DNS queries and responses on a network, likely using a packet analyzer or sniffer such as Wireshark.

Here's a summary of the events:

1. Initial DNS query: The system sends a DNS query for `b'amd.vendors.slashdot.org'` (Packet 113-116).
2. Response: The DNS server responds with IP address `216.34.181.47` (Packet 115-119).
3. Repeated DNS queries: The system repeats the same DNS query (`b'amd.vendors.slashdot.org'`) multiple times, receiving repeated responses of `216.34.181.47`.
4. Repeated DNS queries for other sites: The system sends subsequent DNS queries for other sites:
   * `b'rss.slashdot.org'` (Packet 117-118) -> response with IP address `216.34.181.47`
   * `b'apache.slashdot.org'` (Packet 120) -> response with IP address `216.34.181.48`.
5. ICMP errors: There are multiple ICMP (`dest unreach port unreachable`) packets sent to the system, indicating that it is unable to reach a destination or receive data.
6. Repeated DNS queries for entertainment and games sites: The system sends repeated DNS queries for the `b'entertainment.slashdot.org'` and `b'games.slashdot.org'` sites, receiving responses with IP address `216.34.181.48`.

Some observations:

* The system is repeatedly sending DNS queries, which may indicate a problem with its DNS resolution or a need to fetch data from these websites.
* There are multiple ICMP errors, indicating that the system is experiencing connectivity issues.
* Some responses (e.g., `216.34.181.47`) appear to be repeated, while others (e.g., `216.34.181.48`) have unique IP addresses.

To further investigate this issue, you may want to:

1. Check the system's DNS configuration and ensure that it is correctly configured.
2. Run a network connection test to identify any connectivity issues.
3. Use tools like Wireshark or Tcpdump to capture and analyze network traffic in more detail.
4. Investigate if there are any security vulnerabilities or software issues causing the repeated DNS queries and ICMP errors.

Figure 3: Analysis of network packets from PCAP files.

Enron dataset. The results obtained reflected the typical understanding of a large language model. The amount of data that can be extracted from the Enron email dataset depends highly upon the prompt and the model finetuning. Scout was able was to summarize the emails and raise observations from the chats such as meetings and purchases. Two sample runs shown in figures 4 and 5 Scout has successfully identified

## 4.3 Office Document Analysis

The office documents were first identified and processed using Docling, following which the extracted document layout and content was passed on to Hermes3 model for analysis and summary, while raising any red flags. A sample document analysis is shown in figure 6, with the document containing simple text but manipulated metadata. For the sake of experiments, the document's creation date was changed to a date in the future. The large language model when presented with the document content and metadata quickly recognized this as a red flag along with document editing being done by a user name *Admin*.

## 4.4 Audio File Analysis

To verify the applicability of foundational models in audio evidence files analysis, audio samples from LibreVox project [57] were taken for analysis. OpenAI's Whisper model was used to transcribe the audio into text. Following which the next step involved sending the transcribed text to a large language model for anomaly detection. This XXX

## 4.5 Image Analysis Results

Image analysis is tricky since the larger and clearer the image is, the more accurate the analysis is. There is however, a downside to larger images. The larger an image, the longer it will take to be processed. Image downsizing is a probable avenue that can be explored for an optimal compromise. During the analysis, multiple images were processed. Scout deployed image processing foundational models and was not only able to infer the image contents but also it was able to gauge the context of the images. For analysis, images were taken from the GovDocs1 dataset from Digital Corpora [56]. Figure 7 and 9 were two such images. These images were given to Scout for processing and Scout was able to process and identify the image setting and context with no problems as shown in figure 8 and 10

## 4.6 Video Analysis Results

Video analysis time is directly proportional to the video's size and resolution. Larger and higher resolution videos require more resources. As mentioned earlier Scout uses Qwen2-VL Multimodal model to process video evidence. The model supports video processing of video duration of more than twenty minutes. The depth of inference depends upon the model size. The results from sample video files processing are shown in figures 13 and 14.

# 5 Discussion

Scout has been designed as a pre-investigation platform that allows forensic investigators to identify artifacts that may potentially be of interest. Scout works by sifting through multiple types of files and processing them to identify potential case related content. This often comes in handy when an investigator has to analyze a large collection of files and needs a starting point. Scout provides a starting point wherein the investigator can provide case related information, keywords that can be fed into the framework. The framework can be deployed on a completely offline on premises hardware completely air-gapped from the Internet. Scout functions in read-only mode so that no evidence is ever tampered with. The Scout framework employs artificial intelligence and large language models, to produce the results. Since the large language models are known to hallucinate and be unpredictable any evidence processed by them are inadmissible or in the best case hard to be admissible in court of law. Thus at no point during the whole investigation is Scout involved in actual evidence detection, its extraction or its presentation. Instead Scout points the forensic investigator in a potential important direction. It is then up to the forensic investigator to analyze the evidence files either manually or using time-tested law enforcement approved tools. The results obtained by the investigator using law enforcement approved methods and tools are then documented and submitted as evidences, removing any questions and doubts over the involvement of AI and large language models in the investigation.

Another important fact that the investigator must keep in mind is that apart from false positives, there is a large possibility of false negatives. A false negative occurs whenever the Scout framework processes a file of interest but doesn't raises a flag. This scenario is well understood and is the reason why the Scout framework is just a preliminary tool to identify files that may be of interest. *All the remaining files are must also be analyzed by the forensic investigator to look for any evidences that may have been overlooked.*

# 6 Conclusion & Future Work

This work proposes the application of large language models in the domain of digital forensics to allow a forensic investigator to quickly sift through gigabytes of data and quickly gain insights into potential directions of investigations. This work also presents Scout a prototype framework to achieve the same. Scout is a first step towards the application of large language models in the domain of digital forensics. Large language models have the potential to achieve enormous gains in this field due to their ability to consume large amount of multimodal data process them relatively fast. Future directions in this domain include finetuning large language models to accurately identify and sieve artifacts from vast amount of data files. Finetuning models to specifically cater to the digital forensics community is definitely a future direction that needs extensive work. Specialized models produced can be used to perform course grain and fine grained keyword searches on large corpus of seized data. Larger foundational models can be used for inferring complex data while smaller models can be used for quicker processing of simpler tasks. Finetuned vision models can be used to detect specific images that are of keen interest to the investigation.

Here are the key points from the emails:

1. [illegible] recommends offering two options to New Power Company (NPC) regarding collateral thresholds:
   Option 1: $10MM threshold with standard collateral annex terms
   Option 2: $40MM threshold with standard annex terms plus additional covenants around maintaining minimum liquidity of $300MM and tangible net worth of $450MM

2. [illegible] Buy doesn't like a particular deal but approves it as written.

3. [illegible] had legal problems related to ENEL that [illegible] said he would follow up on.

4. [illegible] made bold moves at [illegible] which surprised [illegible], who planned to call him when in London.

5. [illegible] provided an update on the portfolio management system which [illegible] approved of but requested not to over-engineer the solution.

In summary, the emails cover topics like setting collateral terms for a new client company, providing approvals on deals, following up on legal issues, discussing personnel changes at another firm, and getting updates on internal systems projects. The main themes are financial deal structuring, legal and compliance oversight, and project management within Enron's finance and risk functions.

Figure 4: Sample email Analysis from Enron Dataset (Names blurred).

The email appears to be a communication within Enron, discussing an internal process related to the Peer Review Committee (PRC) and bonuses for the year 2000. The sender, [illegible], is informing the recipients, [illegible] and [illegible], about the requirement to submit suggested reviewers through their system if they want to be considered for a bonus.

Potential Concerns:
1. The tone of the email seems somewhat authoritative and urgent, which might suggest some pressure or importance related to the PRC process.
2. The mention of a "mandatory" process for being considered for a bonus could potentially raise questions about internal policies and fairness.
3. There is no indication of what "Dorothy" refers to in the context, which could be a project, an individual, or something else. This lack of clarity might warrant further investigation.

Metadata Analysis:
- The email was sent on October 31, 2000, at 2:40 AM, which is an unusual time for official company communication. This might suggest that the matter was urgent or that the sender was working irregular hours.
- The email was filed in [illegible]'s folder, indicating that it was likely important to him and possibly related to his work responsibilities.

Overall, while there are some potential concerns and ambiguities in this email, they do not appear to be significant red flags for malicious intent or fraudulent behavior. However, the context and importance of the PRC process might warrant further investigation within the broader scope of Enron's internal operations and policies.

Figure 5: Sample email Analysis from Enron Dataset (Names blurred).

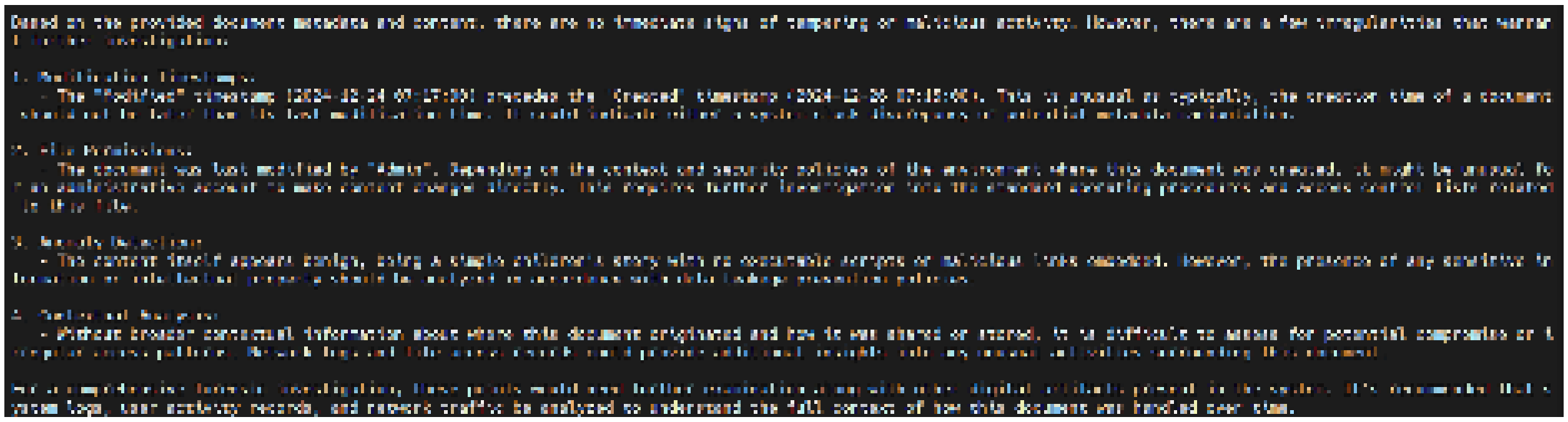

Figure 6: Analysis of a Tampered office document with creation date set later than the last modified date.

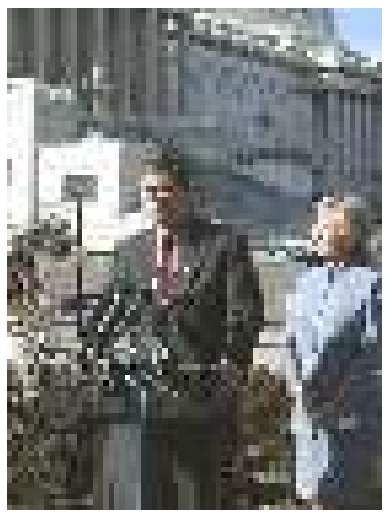

Figure 7: Sample image taken from Digital Corpora [56].

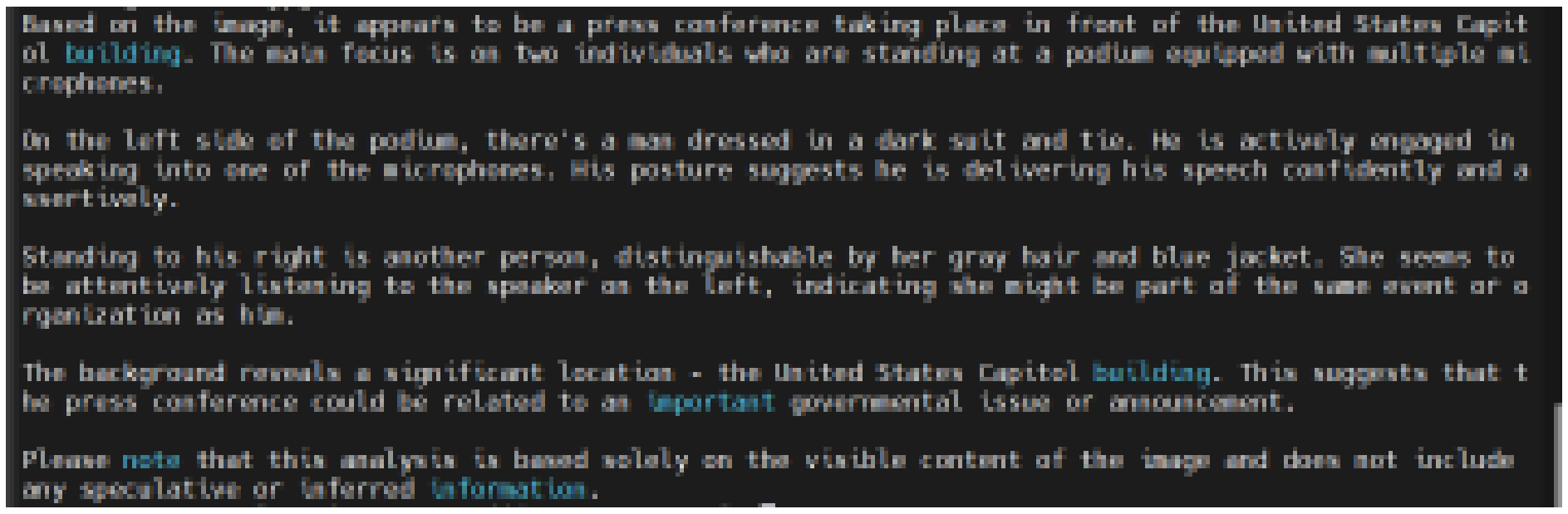

Based on the image, it appears to be a press conference taking place in front of the United States Capitol building. The main focus is on two individuals who are standing at a podium equipped with multiple microphones.

On the left side of the podium, there's a man dressed in a dark suit and tie. He is actively engaged in speaking into one of the microphones. His posture suggests he is delivering his speech confidently and assertively.

Standing to his right is another person, distinguishable by her gray hair and blue jacket. She seems to be attentively listening to the speaker on the left, indicating she might be part of the same event or organization as him.

The background reveals a significant location - the United States Capitol building. This suggests that the press conference could be related to an important governmental issue or announcement.

Please note that this analysis is based solely on the visible content of the image and does not include any speculative or inferred information.

Figure 8: Information gathered from the previous image. The model is able to identify the backdrop of U.S. Capitol Building.

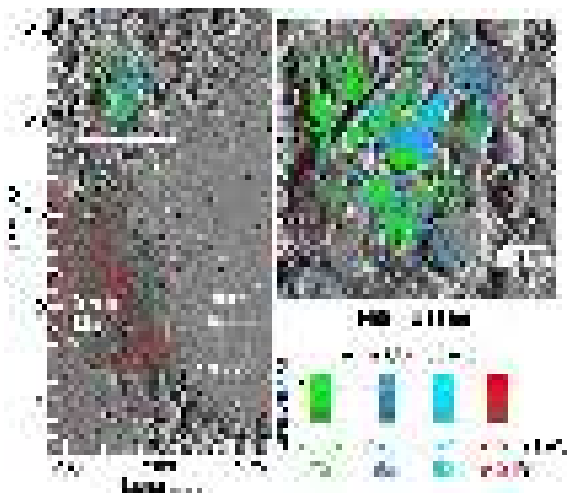

Figure 9: Sample image taken from Digital Corpora [56].

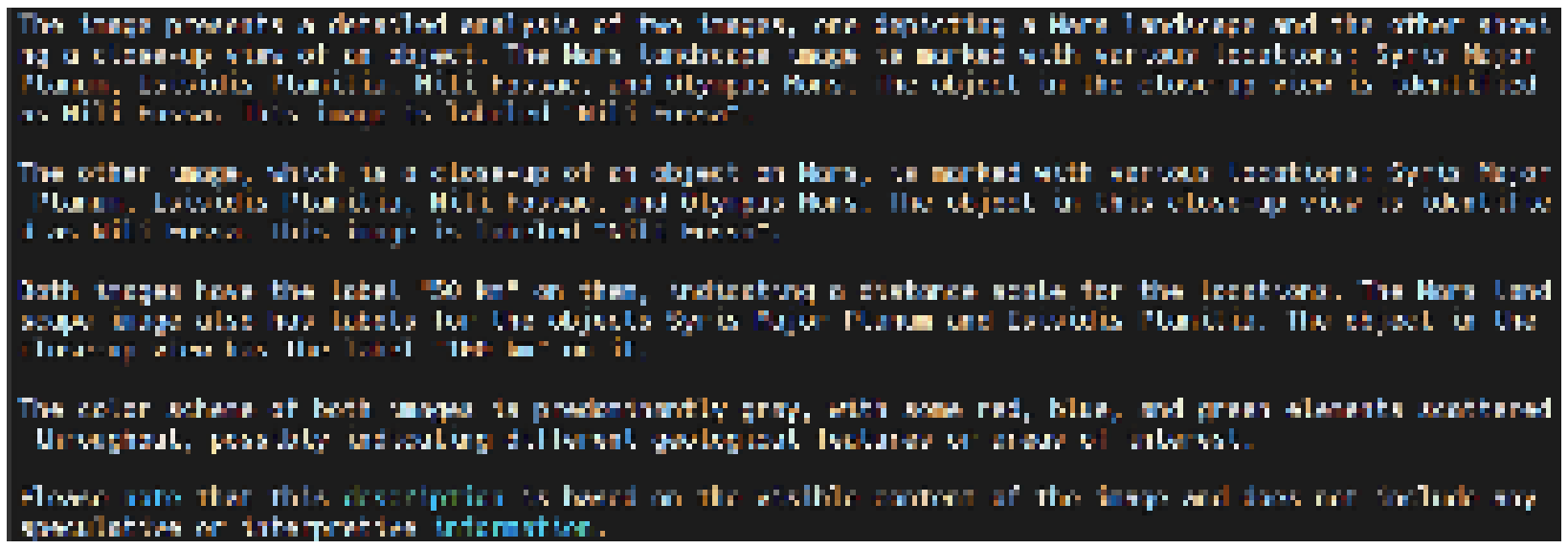

The image presents a detailed analysis of two images, one depicting a Mars landscape and the other showing a close-up view of an object. The Mars landscape image is marked with various locations: Syrtis Major Planum, [illegible] Planitia, Nili Fossae, and Olympus Mons. The object in the close-up view is identified as Nili Fossae. This image is labeled "Nili Fossae".

The other image, which is a close-up of an object on Mars, is marked with various locations: Syrtis Major Planum, [illegible] Planitia, Nili Fossae, and Olympus Mons. The object in this close-up view is identified as Nili Fossae. This image is labeled "Nili Fossae".

Both images have the label "50 km" on them, indicating a distance scale for the locations. The Mars landscape image also has labels for the objects Syrtis Major Planum and [illegible] Planitia. The object in the close-up view has the label "100 km" on it.

The color scheme of both images is predominantly gray, with some red, blue, and green elements scattered throughout, possibly indicating different geological features or areas of interest.

Please note that this description is based on the visible content of the image and does not include any speculative or interpretive information.

Figure 10: Information gathered from the previous image. The model is able to identify the planet Mars landscape in the image.

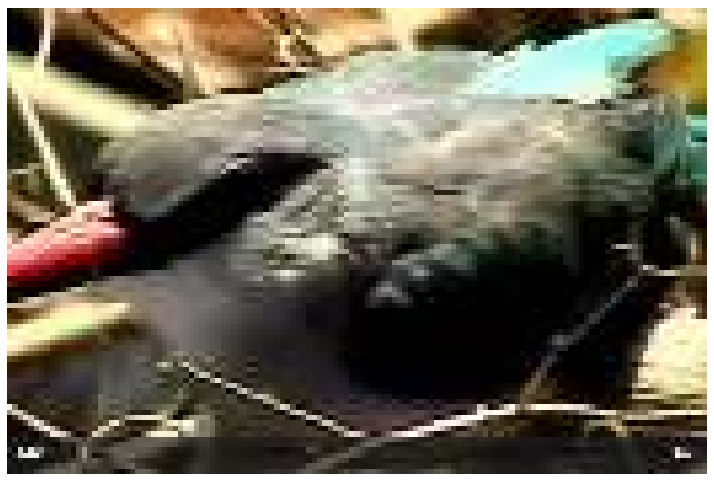

Figure 11: Sample video taken from Digital Corpora [56].

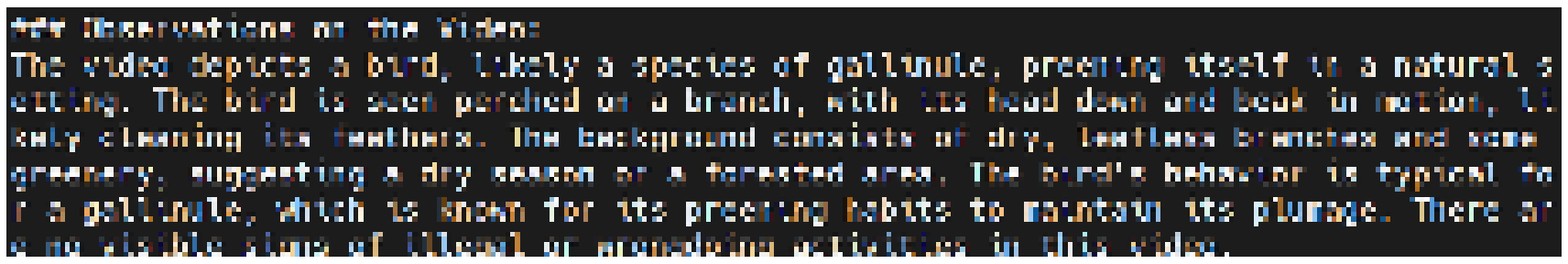

### Observations on the Video:
The video depicts a bird, likely a species of gallinule, preening itself in a natural setting. The bird is seen perched on a branch, with its head down and beak in motion, likely cleaning its feathers. The background consists of dry, leafless branches and some greenery, suggesting a dry season or a forested area. The bird's behavior is typical for a gallinule, which is known for its preening habits to maintain its plumage. There ar[illegible]

Figure 12: Information gathered from the previous image. The model is able to identify the birds activity as well as species along with the backdrop.

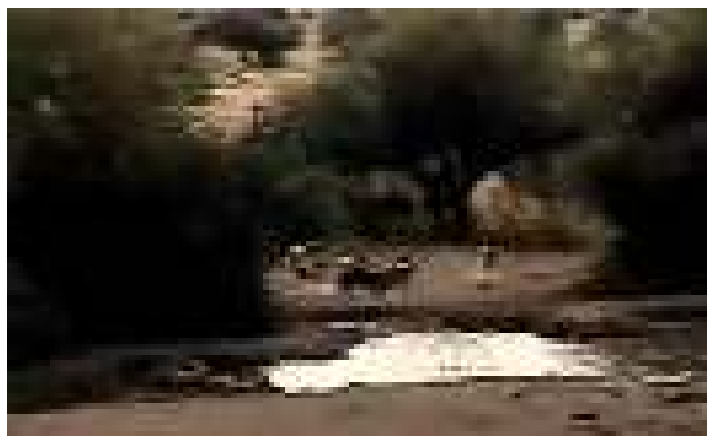

Figure 13: Sample video taken from Digital Corpora [56].

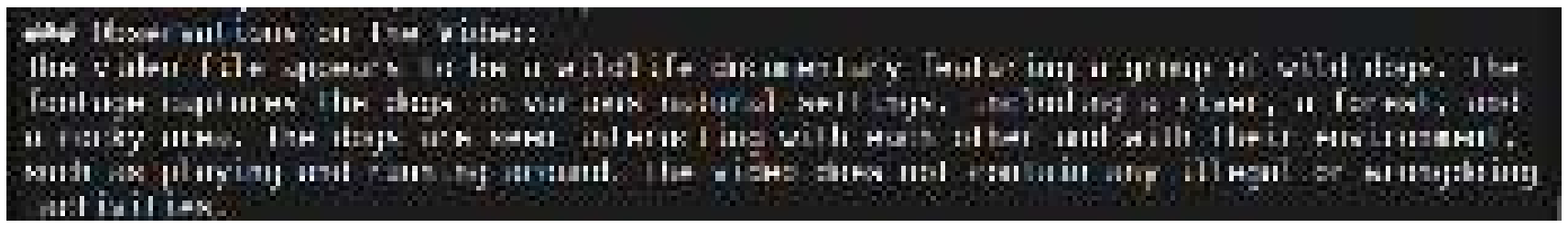

### Observations on the Video:
The video file appears to be a wildlife documentary featuring a group of wild dogs. The footage captures the dogs in various natural settings, including a river, a forest, and a rocky area. The dogs are seen interacting with each other and with their environment, such as playing and running around. The video does not contain any illegal or [illegible] activities.

Figure 14: Information gathered from the previous image. The model is able to identify the animals (wild dogs) and their activities along with their environmental backdrop along with a comment on the activities.